\documentstyle[12pt]{article}

\def\thebibliography#1{\section*{
References}\list
  {\arabic{enumi}.}{\settowidth\labelwidth{#1}\leftmargin\labelwidth
    \advance\leftmargin\labelsep
    \usecounter{enumi}}
    \def\newblock{\hskip .11em plus .33em minus .07em}
    \sloppy\clubpenalty4000\widowpenalty4000
    \sfcode`\.=1000\relax}

\def\op#1{\mathop{\fam0 #1}\limits}

\newcommand{\beq}{\begin{equation}}
\newcommand{\eeq}{\end{equation}}
\newcommand{\ben}{\begin{eqnarray}}
\newcommand{\een}{\end{eqnarray}}
\newcommand{\be}{\begin{eqnarray*}}
\newcommand{\ee}{\end{eqnarray*}}
\newcommand{\bea}{\begin{eqalph}}
\newcommand{\eea}{\end{eqalph}}

\newcommand{\cL}{{\cal L}}
\newcommand{\cH}{{\cal H}}
\newcommand{\bR}{{\bf R}}
\newcommand{\al}{\alpha}
\newcommand{\bt}{\beta}

\newcommand{\la}{\lambda}
\newcommand{\f}{\phi}
\newcommand{\om}{\omega}
\newcommand{\Om}{\Omega}
\newcommand{\m}{\mu}
\newcommand{\th}{\theta}

\newcommand{\si}{\sigma}
\newcommand{\w}{\wedge}
\newcommand{\wh}{\widehat}
\newcommand{\ol}{\overline}
\newcommand{\dr}{\partial}
\newcommand{\ot}{\otimes}

\newenvironment{eqalph}{\stepcounter{equation}
\setcounter{equationa}{\value{equation}}
\setcounter{equation}{0}

\begin{eqnarray}}{\end{eqnarray}
\setcounter{equation}{\value{equationa}}}

\begin{document}
\hbox{}

\begin{center}
{\large \bf ON THE BRACKET PROBLEM IN COVARIANT  
\medskip 

HAMILTONIAN FIELD THEORY}
\bigskip

{\sc L.Mangiarotti}

Department of Mathematics and Physics, Camerino University, 62032 Camerino,
Italy

E-mail: mangiaro@camserv.unicam.it
\medskip

{\sc G.Sardanashvily}

Department of Theoretical Physics, Moscow State University

117234 Moscow, Russia

E-mail: sard@grav.phys.msu.su 
\end{center}

\bigskip

\begin{abstract}
The polysymplectic phase space of covariant Hamiltonian field theory can be
provided with the current algebra bracket.
\end{abstract}
\bigskip

 As is well known, when applied to field theory, the familiar
symplectic technique of mechanics takes the form of instantaneous Hamiltonian
formalism on an infinite-dimensional phase space. 
The finite-dimensional covariant Hamiltonian approach to field theory is
vigorously developed from the seventies in its multisymplectic and
polysymplectic variants (see \cite{book,got91,got92,sard95} for a
survey). They are related to the two different Legendre morphisms in the first
order calculus of variations on fiber bundles.

Recall that, given a fiber bundle
$Y\to X$, coordinated by $(x^\la,y^i)$, a first order Lagrangian $L$ is
defined as a semibasic  density
\be
L=\cL\om: J^1Y\to\op\w^nT^*X, \quad \om=dx^1\w\cdots dx^n, \quad n=\dim X,
\ee
on the affine jet bundle $J^1Y\to Y$, provided with the adapted coordinates
$(x^\la,y^i,y^i_\la)$. $J^1Y$ can be seen as a  finite-dimensional
configuration space of fields represented by sections of $Y\to X$. 
The Poincar\'e--Cartan form
\be
 H_L=\cL\om +\pi^\la_i(dy^i-y^i_\m dx^\m)\w\om_\la, \quad
\pi^\la_i=\dr^\la_i\cL,
\quad
\om_\la=\dr_\la\rfloor\om, 
\ee
can be defined as a Lepagean equivalent of $L$ which is a semibasic form on
$J^1Y\to Y$. It yields the Legendre bundle morphism
$\wh H_L$ over $Y$ of
$J^1Y$ to the homogeneous Legendre bundle
\beq
Z_Y=J^{1\star}Y = T^*Y\w(\op\w^{n-1}T^*X)  \label{N41}
\eeq
which is the affine $\op\w^{n+1}$-valued dual of $J^1Y\to Y$, and is endowed
with the holonomic coordinates $(x^\la,y^i,p^\la_i,p)$. 
In these coordinates, the morphism (\ref{N41}) reads
\be
(p^\m_i, p)\circ\wh H_L =(\pi^\m_i, \cL-\pi^\m_i y^i_\m). 
\ee
The fiber bundle $Z_Y$ is treated as a homogeneous finite-dimensional phase
space of fields. It is equipped with the canonical multisymplectic form
\be
d\Xi_Y= dp\w\om + dp^\la_i\w dy^i\w\om_\la. 
\ee

Every Lagrangian $L$ defines the Legendre map $\wh L$ over
$Y$ of $J^1Y$ to the Legendre bundle 
\beq
\Pi=\op\w^nT^*X\op\ot_YV^*Y\op\ot_YTX \label{00}
\eeq
provided with the holonomic coordinates $(x^\la,y^i,p^\la_i)$. $\Pi$
plays the role of a finite-dimensional momentum phase space of fields. It is
equipped with the canonical polysymplectic form
\be
\Om_Y =dp_i^\la\w dy^i\w \om\ot\dr_\la. 
\ee

The
relationship  between the multisymplectic and polysymplectic phase spaces is
given by the exact sequence
\be
0\to\Pi\op\times_X\op\w^nT^*X\hookrightarrow Z_Y\to\Pi\to 0, 
\ee
where $\pi_{Z\Pi}:Z_Y\to \Pi$ is a 1-dimensional affine bundle. Given a
section
$h$ of $Z_Y\to\Pi$, the pull-back $h^*\Xi_Y$ is a
polysymplectic Hamiltonian form on $\Pi$
\cite{cari,book,sard95}. 

A natural idea is to generalize a Poisson
bracket in symplectic mechanics to multisymplectic or polysymplectic
manifolds and, as a final result, to come to the covariant canonical
quantization of field theory. Different variants of such a bracket have
been suggested (see \cite{kan1,kan2} and references therein). The main
difficulty is that the bracket must be globally defined. 
Let us note that multisymplectic manifolds, whose particular example is
$(Z_Y,d\Xi_Y)$, look rather promising for algebraic constructions since
multisymplectic forms are exterior forms \cite{cantr,cantr1,hrab,ibort,law}.  

Nevertheless, the $X=\bR$ reduction of the covariant
Hamiltonian formalism leads to time-dependent mechanics, but not conservative
symplectic mechanics \cite{book98,sard98}.
In this case, $Z_Y=T^*Y$ and $\Pi=V^*Y$ are homogeneous and momentum phase
spaces of time-dependent mechanics, respectively. The momentum phase space
$V^*Y$, coordinated by $(t,y^i,p_i=\dot y_i)$, is endowed with the canonical
degenerate Poisson structure given by the Poisson bracket
\beq
\{f,g\}_V=\dr^if\dr_i g-\dr^ig\dr_i f, \qquad f,g\in C^\infty(V^*Y).
\label{c3}
\eeq
However, a Poisson bracket $\{\cH,f\}_V$ of a Hamiltonian $\cH$ and
functions $f$ on the momentum phase space $V^*Y$ fails to be a well-behaved
entity because a Hamiltonian of time-dependent mechanics is not a scalar with
respect to time-dependent transformations, e.g., time-dependent canonical
transformations. In particular, the equality $\{\cH,f\}_V=0$ is not preserved
under time-dependent transformations. As a consequence, the evolution
equation in time-dependent mechanics is not reduced to a Poisson bracket, and
integrals of motion are not functions in involution with a Hamiltonian
\cite{book98,sard98}.

At the same time, the Poisson bracket (\ref{c3}) leads to the following
current algebra bracket. Let
$u=u^i\dr_i$ be a vertical vector field on $Y\to\bR$, and $J_u=u^ip_i$ the
corresponding symmetry current on $V^*Y$ along $u$. In particular, $J_u$ is a
conserved N\"other current if a Hamiltonian form 
\be
H=p_idy^i-\cH dt
\ee
is invariant under the 1-parameter group of local gauge transformations whose
generator is the vector field $u$ \cite{book98,sard98}. The symmetry currents
$J_u$ constitute a Lie algebra with respect to the bracket
\be
[J_u,J_{u'}]=\{J_u,J_{u'}\}_V= J_{[u,u']}.
\ee
This current algebra bracket can be extended to the general polysymplectic case
as follows.

There is the canonical isomorphism
\be
\ol\th= p^\la_i\ol dy^i\w\om_\la: \Pi\to
V^*Y\op\w_Y(\op\w^{n-1}T^*X),
\ee
where $\{\ol dy^i\}$ are the fiber bases for the vertical cotangent
bundle $V^*Y$ of $Y\to X$. Let $u=u^i\dr_i$ be a vertical vector field on
$Y\to X$. The corresponding symmetry current is defined as a
semibasic exterior $(n-1)$-form
\beq
J_u=u\rfloor\ol\th=u^ip_i^\la\om_\la \label{c1}
\eeq
on the Legendre bundle $\Pi$ (\ref{00}). In particular, $J_u$ is a  N\"other
current if a Hamiltonian form $H$ is invariant under the 1-parameter group of
local gauge transformations whose generator is the lift onto $\Pi$ of the
vector field $u$
\cite{book,sard97}. The symmetry currents (\ref{c1}) constitute a Lie algebra
with respect to the bracket
\beq
[J_u,J_{u'}]\op=^{\rm def} J_{[u,u']}. \label{c2}
\eeq
If $Y\to X$ is a vector bundle and $X$ is provided with a non-degenerate
metric $g$, the bracket (\ref{c2}) can be extended to any semibasic
exterior $(n-1)$-forms $\f=\f^\al\om_\al$ on $\Pi$ by the law
\be
[\f,\si]=g_{\al\bt}g^{\m\nu}(\dr^i_\m\f^\al\dr_i\si^\bt-
\dr^i_\m\si^\bt\dr_i\f^\al)\om_\nu.
\ee
Similarly, the bracket of semibasic 1-forms on $\Pi$ is defined \cite{book98}.

The bracket (\ref{c2}) looks promising for the current algebra quantization
of the covariant Hamiltonian formalism.

\end{document}